\documentclass[prx,aps,showpacs,twocolumn,preprintnumbers,
amsmath,amssymb,superscriptaddress,longbibliography,nofootinbib]{revtex4-1}

\usepackage{graphicx}
\usepackage{dcolumn}
\usepackage{bm}
\usepackage{color,soul}

\begin{document}
\newcommand{\iqoqi}{Institute for Quantum Optics and Quantum Information IQOQI, Boltzmanngasse 3, 1090 Vienna, Austria.}
\newcommand{\univie}{University of Vienna - Faculty of Physics, Boltzmanngasse 5, 1090 Vienna, Austria.}
\newcommand{\qtlabs}{Quantum Technology Laboratories, Clemens-Holzmeister-Straße 6/6, 1100 Vienna, Austria.}
\newcommand{\mpl}{Max Planck Institute for the Science of Light, Staudtstrasse 2, 91058 Erlangen, Germany.}

\title{Quantum interference between distant creation processes}
    \author{Johannes Pseiner}
        \email{johannes.pseiner@qtlabs.at}
        \affiliation{\qtlabs}%
    \author{Manuel Erhard}%
        \email{manuel.erhard@qtlabs.at}
        \affiliation{\univie}%
        \affiliation{\iqoqi}%
    \author{Mario Krenn}%
        \email{mario.krenn@mpl.mpg.de}
        \affiliation{\mpl}%
    \date{\today}

\begin{abstract}
The search for macroscopic quantum phenomena is a fundamental pursuit in quantum mechanics. It allows us to test the limits quantum physics and provides new avenues for exploring the interplay between quantum mechanics and relativity. In this work, we introduce a novel approach to generate macroscopic quantum systems by demonstrating that the creation process of a quantum system can span a macroscopic distance. Specifically, we generate photon pairs in a coherent superposition of two origins separated by up to 70 meters. This new approach not only provides an exciting opportunity for foundational experiments in quantum physics, but also has practical applications for high-precision measurements of distributed properties such as pressure and humidity of air or gases.
    
\end{abstract}

\maketitle

\section{\label{sec:level1}INTRODUCTION}
\begin{figure}
\centering
\includegraphics[width=3.21in]{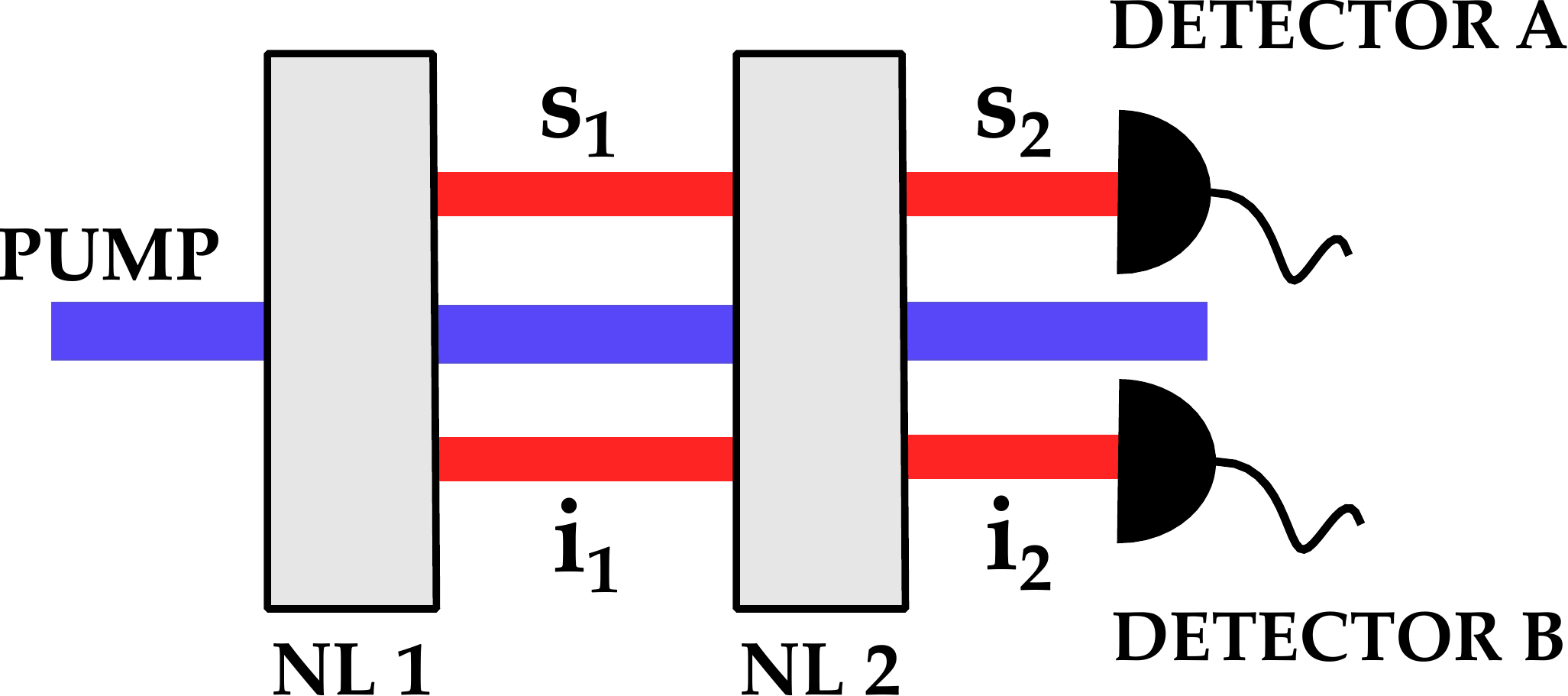}
\caption{\small{A schematic picture of the simplified experimental arrangement is depicted. A continuous-wave pump laser was used to create a down-converted photon pair in a nonlinear crystal in the modes $s_{1}$ and $i_{1} .$ Within the same modes the pump beam propagated collinearly to the second crystal so an additional possibility for creating photon pairs indicated by $s_{2}$ and $i_{2}$ arises. Aligning both signal and idler beams such that the which-crystal information is removed, interference fringes can be observed while scanning the phase difference between the pump and down-conversion beams. These effects shall be observed with increased spatial separation of the crystals. Taken from \cite{pseinerdiss}.}}
\label{fig:conceptsketch}
\end{figure}

In quantum mechanics, if two alternatives cannot be distinguished -- even in principle -- interference can occur. Feynman said that this property "has in it the heart of quantum mechanics" \cite{feynman1965feynman}. In 1994, Herzog et al. \cite{herzog1994frustrated} demonstrated that this phenomenon can not only be observed for properties of individual or entangled photons, but for the creation process of photons themselves. Expanding an experiment by Zou, Wang and Mandel \cite{zou1991induced}, they have overlapped the paths of a photon pair generated by one creation process with the paths generated by another creation process in such a way. The setup, depicted in Fig. \ref{fig:conceptsketch} has been aligned in such a way that there is no information (not even in principle) to find out in which of the two creation process the photon pair has been generated. Therefore the photon pair is in a coherent position of being created in the first or second process. By adapting a phase between the two processes, constructive and destructive interference can be observed. For constructive interference, the total number of generated photon pairs can be enhanced by a factor of four compared to a single crystal, while for destructive interference, the number of generated photon pairs is zero. A conceptual sketch of this experiment can be seen in Fig. \ref{fig:conceptsketch}.

This peculiar quantum phenomenon has been employed in recent years for numerous applications \cite{hochrainer2022quantum}, ranging from spectroscopy \cite{kalashnikov2016infrared} to sensing \cite{kutas2020terahertz} and entanglement generation \cite{hardy1992source, krenn2017entanglement}. So far, this process has only been observed with very small spatial distance between the two creation processes. Either, the two creation processes occur at the same location (for instance, by pumping the same nonlinear crystal from two directions \cite{herzog1994frustrated, qian2021multiphoton}), or are separated at the millimeter scale (for instance, at an integrated photonic chip \cite{ono2019observation, feng2023chip}).

Here, we observe coherent quantum superposition between the origins of two macroscopically separated photon creation processes. Specifically, we use two nonlinear crystal, spatially separated by up to 70 meters. Each crystal can create photon pairs. Importantly, by overlapping the paths of the photons from the two crystals, we create a scenario in which the generated photon pairs cannot reveal any distinguishing information about their origin.

Pushing the distance between two creation processes has multiple motivations. First, on a technical side, one can envision highly sensitive, quantum enhanced sensing methods for large scale properties such as air pressure or temperature fluctuations. Second, spatially separating the creation process is necessary for demonstrating the nonlocal nature of a new multiphoton quantum interference effects without explicit entanglement, which has been theoretically proposed in \cite{gu2019quantum, hochrainer2022quantum} and observed at a local scale in \cite{feng2023chip, qian2021multiphoton}. Third, at a fundamental physics level, the unusual feature of our experiment is that the entire generation process of the quantum state scales over a macroscopic distance. Thus, our experiment is a first step towards a new way of testing the limits of ever larger and more complex quantum systems and the concept of \textit{macroscopicity} \cite{leggett2002testing}, complementary to quantum systems with large masses \cite{fein2019quantum}, photon numbers \cite{bruno2013displacement}, angular momentum \cite{fickler2012quantum} or large scale entanglement distribution \cite{yin2017satellite}.

\section{METHODS}
\label{sec:methods}

In our work, we choose two interfering quantum processes as an experimental demonstration for macroscopically large quantum systems. The idea is simple: Investigate how much we can expand a seemingly coherent quantum system without losing the associated quantum effects. Our system is comprised of two quantum processes that create pairs of photons in a nonlinear optical process called spontaneous parametric down-conversion (SPDC). The reason for choosing such a system is manifold. It can be used for long-distance experiments and multi-photon experiments. Hence it is extendable in terms of space, the number of photons, and the dimensionality of information contained in the system. The principle of path identity allows for a conceptually simple implementation of the proposed idea. The SPDC process can be approximated by a power series expansion \cite{walborn2010spatial}
\begin{figure}
\centering
\includegraphics[width=3.21in]{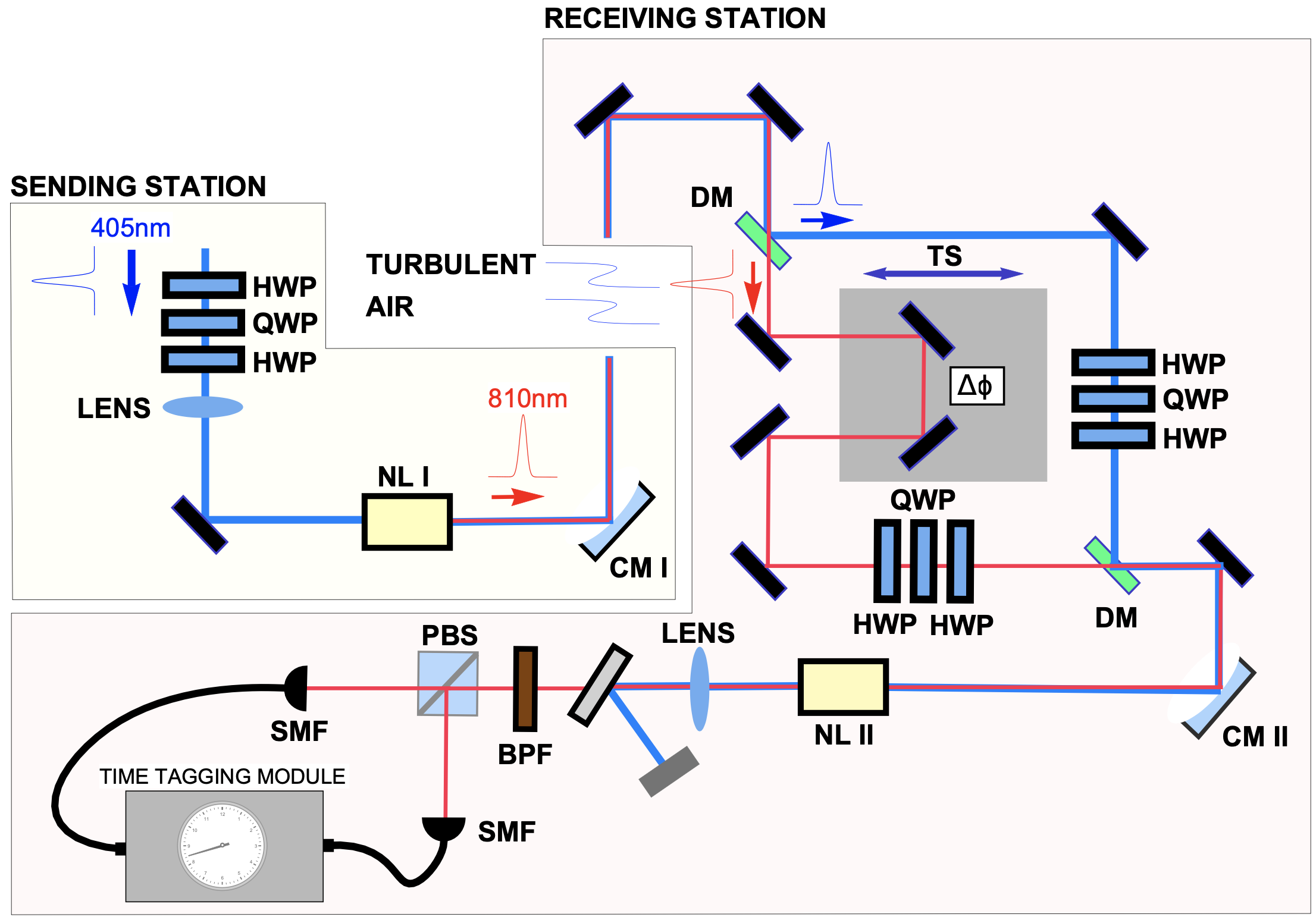}
\caption{\small{A schematic picture of the experimental setup, which contains the coherent pumping of two nonlinear crystals (NL I at the sending station and NL II at the receiving station) with pump and down-converted beams propagating collinearly, is shown. The phase difference $\Delta \phi$ was introduced via a trombone system (TS) within a Mach-Zehnder interferometer after splitting up the pump and the down- converted beam and combining them again with a dichroic mirror (DM). To avoid chromatic aberration of the lenses between pump and SPDC photons, two concave mirrors (CM I and II) for both sending and receiving the signals were used. To filter out the undesired pump signal as well as to narrow down the wavelength distribution, bandpass filters (BPF) were implemented in the detection system. The recorded detection events were labeled with a timestamp provided by a time tagging module. Simultaneous clicks within a coincident timing window tc, which was chosen to be 1.5ns, were identified as coincidences. Taken from \cite{pseinerdiss}.}}
\label{fig:full_setup}
\end{figure}
\begin{equation}
S_{ab}=1+g_{ab}(a^\dagger b^\dagger ) + g_{ab}^2/2(a^\dagger b^\dagger )^2+O(g_{ab}^3),	
\end{equation}
where the generation rate $g$ is proportional to the second-order nonlinear coefficient and the pump power, and $a^\dagger,b^\dagger$ refer to the creation operator of photons in the paths $a$ and $b$. We neglect all higher orders ($\geq O(g^2)$) for further discussions by operating our experiment at low pump powers, see Supplementary for details.

Inserting another nonlinear crystal into the path of the first according to the path identity principle with a general phase $U_{\phi}$ between the two processes leads to 
\begin{equation}
    \begin{split}
	S_{cd} U_{\phi} S_{ab} = [1+g_{cd}(c^\dagger d^\dagger )] [1+e^{i\phi}g_{ab}(a^\dagger b^\dagger )]\\\nonumber
	=1+g_{cd}c^\dagger d^\dagger+e^{i\phi}g_{ab}a^\dagger b^\dagger+O(g^2),
\end{split}
\end{equation}

    \begin{figure*}
        \centering
            \includegraphics[width=6.6in]{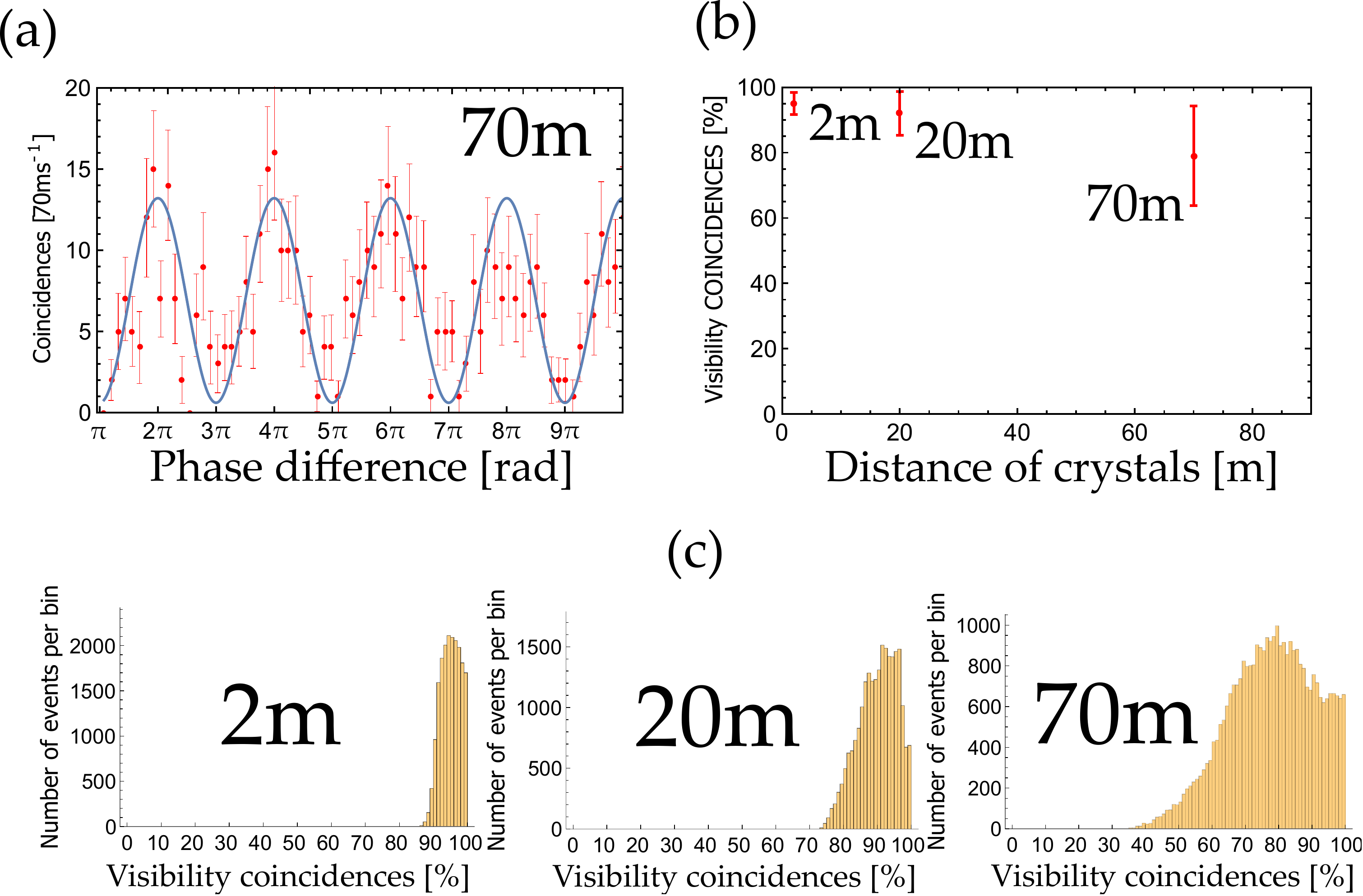}
            \caption{\small{The results of the quantum-interference experiment are depicted. 
            (a) The coincident count rates of down-conversion photons while moving the trombone system and hence changing $\Delta \mathcal{L}$ after a propagation distance of $70\,$m is shown.
            (b) The scaling behavior of the system with increasing distance between nonlinear crystals are shown in terms of the visibility. The red dots correspond to the peak value of the respective distribution and the red error bars equal the standard deviation of the estimated distributions.
            (c) A summary of all distributions of visibilities over the respective propagation distances is shown. 
            }}
        \label{fig:all}
    \end{figure*}

with $\phi$ denoting a phase between the two nonlinear processes. Finally, using the principle of path identity (i.e. overlaying the paths of the single photons such that $c\rightarrow a$ and $d\rightarrow b$), we arrive at
\begin{equation}
\begin{split}
    S_{cd} U_{\phi} S_{ab}|vac\rangle\rightarrow[1+g_{cd}c^\dagger d^\dagger+e^{i\phi}g_{ab}a^\dagger b^\dagger]|vac\rangle\\\nonumber\rightarrow|0,0\rangle_{a,b}+g_{ab}(1+e^{i\phi})|1,1\rangle_{a,b},
\end{split}
\end{equation}
where in the first step the quantum system $S_{cd} U_{\phi} S_{ab}$ acts on the vacuum mode $|vac\rangle$, with $|1,1\rangle_{a,b}$ and $|0,0\rangle_{a,b}$ denoting one or no photon pair in the paths $a$ and $b$, respectively.

Therefore, the derived equation shows that our proposed system either produces pairs of photons or not, depending only on the relative phase $\phi$ between the two processes. Most importantly, there is no theoretical evidence that the spatial distance between the two processes reduces the quantum behavior of the suppressed or enhanced emission of photons. The suppressed and enhanced photon pair emission also shows the fundamental difference between our system and other spatially extended interference or quantum experiments. Our entire setup consists of two spatially distant quantum processes, which, dependent on their relative phase, either produce pairs of photons or not. In this case, it is not the product of a process measured at a large distance from one another, as is the case, for example, with loophole-free bell experiments \cite{larsson2014loopholes,hensen2015loophole,giustina2015significant,shalm2015strong}. Instead, the processes are separated far from one another such that the quantum system extends over a large distance. Theoretically, there is no upper limit to the distance between the two creation processes.

From an experimental perspective, the question arises what does the coherence of the two processes depend on physically? Here, two conditions must be met. The first condition, which is analogous to Eq. (6), is that the optical pathlength difference between the pump beam and the two down-converted photons must be smaller than the coherence length of the pump laser,
\begin{equation}
	|L_p-L_{DC}^a-L_{DC}^b|\leq L_p^{coh-len},
\end{equation}	
where $L_p$ denotes the path length of the pump laser, $L_{DC}^{a,b}$ describes the path length of the down-converted photon from the process a or b, and $L_{p}^{coh-len}$ the coherence length of the pump laser.

The second condition is given by the following optical pathlength difference of the down-conversion photons and their coherence length: 
\begin{equation}
	|L_{DC}^a-L_{DC}^b|\leq L_{DC}^{coh-len},
\end{equation}
with $L_{DC}^{coh-len}$ the coherence length of the down-converted photons.

Additionally, all degrees of freedom of the down-converted photons must be identical to achieve perfect indistinguishability. Hence, to ensure path indistinguishability, we overlay the down-converted photons from the two processes and remove residual path information by coupling them into single-mode optical fibers. To ensure identical spectral properties, we utilize narrow-band optical band path filters. We use a quarter-half-quarter (QHQ) waveplate combination to align the polarisations of both down-converted photon pairs. Finally, we control the brightness of both photon creation processes by altering the polarisation of the pump beam before the second nonlinear crystal.

We split the down-conversion and pump beam using dichroic mirrors and introduce path length differences with a trombone system to change the phase between the two quantum processes.

\section{RESULTS}
We confirm the quantumness of the system by measuring the interference between the two distant quantum pair creation processes. The visibility of the interference is defined as $V=(max-min)/(max+min)$, where $max$ describes the two-photon count rate at phase setting $\phi=0$, and $min$ the two-photon count rate $\phi=\pi$.

For a classical, incoherent system, we would expect a vanishing visibility. For a perfect quantum system, we expect to observe interference patterns in the two-photon coincidence events by varying the relative phase between the pump and down-converted photons, according to equation (3). In principle, measuring 2 phase settings ($\phi=0$ and $\phi=\pi$) over a sufficiently long time interval increases the statistical significance of the visibility measurement. However, we chose to alternate the relative phase $\phi$ at a speed of 180nm/second to minimize phase fluctuations from turbulent air and ensure the experiment's proper functioning at all times. Proper functioning here refers to a "null result," meaning no counts detected, which would always yield perfect visibility and could also occur if no photons were produced, as would be the case for a malfunctioning experiment, for example.

Figure 3 shows the observed interference pattern of two quantum processes separated by 70m. The visibility pattern is visible, and we added a sinusoidal curve to guide the eye. The visibility was calculated by extracting the maximas and minimas over a total period of 70 seconds. This resulted in approximately 70 measurement points per minima and maxima setting. Calculating the average visibility from those measurement points results in $83\%\pm15\%$. This main result demonstrates the interference of two spatially distant quantum processes.

In addition, we also show two visibility measurements at shorter distances of 2m and 20 meters, respectively. The average visibilities are $96\%\pm 3\%$ for 2 meters and $92\%\pm 6\%$ for a 20-meter distance. The apparent drop in visibilities with increasing distance between the two quantum processes can be explained in the following way. Due to the atmospheric turbulences caused by the air conditioning system in the laboratory, the pump beam (at 400nm wavelength) experienced an angle of arriving fluctuations at the second nonlinear crystal. These angles of arriving fluctuations resulted in a high variance of created photon pairs and hence different amplitudes of the two-photon creation processes. The different amplitudes effectively lead to decreasing visibility, analogously to standard single photon interference experiments, see Supplementary for details.

\section{DISCUSSION}

    
    A summary of the visibilities of the coincident count rates in dependency of the distance between the crystals is displayed in Fig. \ref{fig:all}.
    While the width of distribution for coincident counts was relatively narrow for $2$m, for longer distances the error bars increased, and the position of the peak went to lower visibilities. 
    Note that due to the high number of evaluated statistics, i.e. high number of photons $n$, during measuring ($70\,$s total measurement time per measurement session) and the accumulated high sampling number in post-processing, the error of the mean value was negligibly small.
    The broadening of the error of the visibility distributions could be explained with the fact that, as for the three measurements the measurement times were equal, the count rates over $70\,$m were lower in magnitude (few counts per integration time $t_{int}=70\,$ms, see Fig. \ref{fig:coinc70m}) compared to the ones for the other distances ($\sim \!10^2$ counts per $t_{int}$, see Fig. \ref{fig:coinc2m} and Fig. \ref{fig:coinc20m}).
    Obviously, higher statistical significance could have been achieved by focusing on the single count rates, which, however, proved to be more experimentally challenging, as the visibility of single count rates was highly dependent on the information about the partner photon.
    By definition, within the coincident count rates, no information was leaked to the environment and coherence was conserved. 

    \section{OUTLOOK}
    In our experiment we show that the birth place of a quantum state can be spatially spread over a distance of 70 meters. With linear interpolations, we expect that visibility above 50\% will be possible up to 250 meters (10\% for 500 meters), without additional control mechanisms. An interesting future research question is how effects from special or general relativity affect the interference in these systems, for example by dephasing or decoherence, as investigated in related proposals \cite{joshi2018space}.

    \begin{figure}
        \centering
            \includegraphics[width=2.91in]{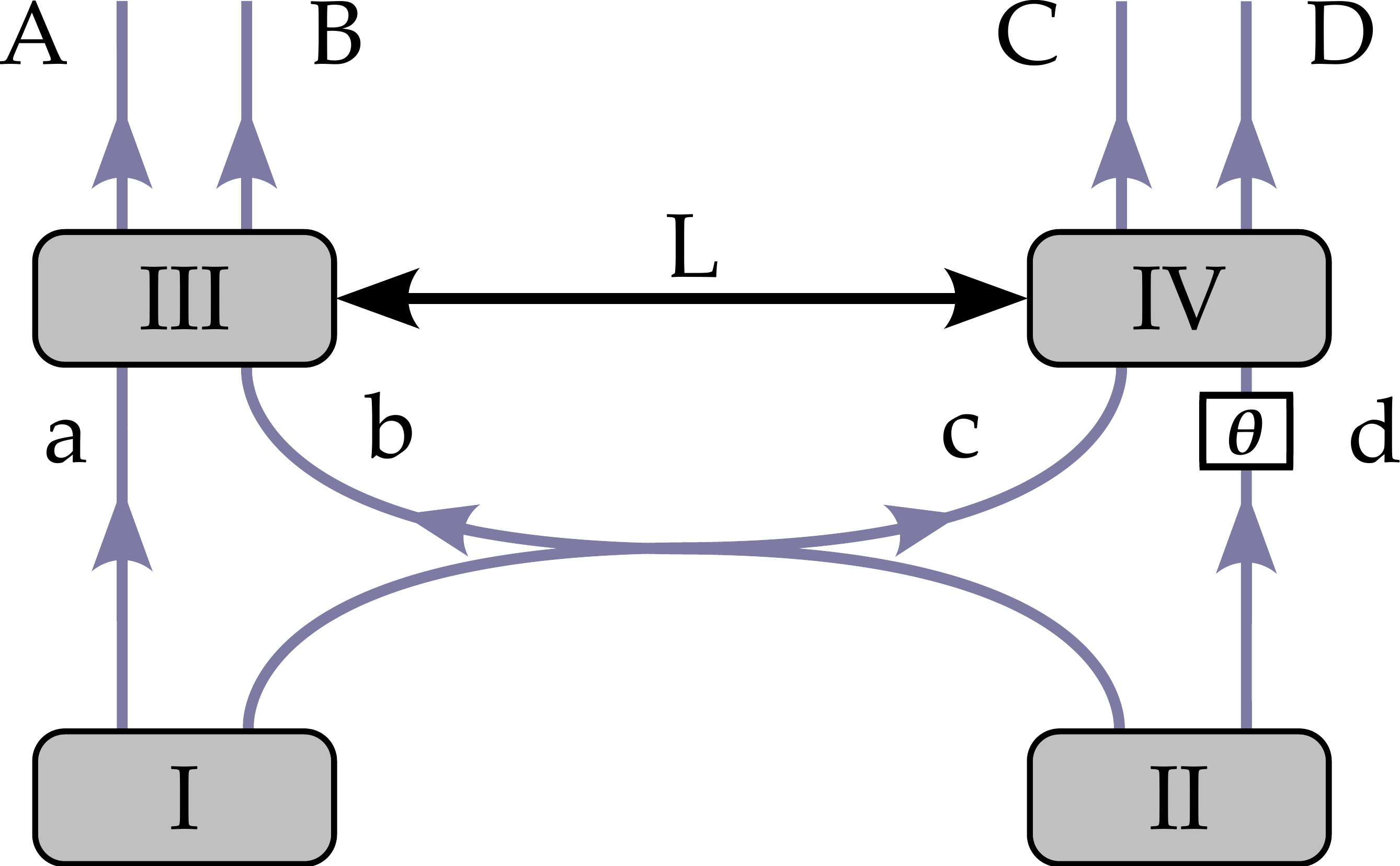}
            \caption{\small{Proposed experimental setup for observing a nonlocal quantum interference phenomenon of the origins of a photon-quatruple. Here, four photons are created either in crystal I$\&$III or II$\&$IV, with a large distance L between the crystals which could lead to the observation of the nonlocal features of a multipartite quantum system. The experiment has been proposed in \cite{gu2019quantum,hochrainer2022quantum} and recently demonstrated for small L in \cite{feng2023chip, qian2021multiphoton}.
            }}
        \label{fig:horizontal_interference}
    \end{figure}
    
    The results presented here are a crucial first step for the observation of new non-local multi-photon interference effects, proposed in \cite{gu2019quantum,hochrainer2022quantum}. There -- similar to the experiment demonstrated here -- a four-photon quantum state is generated in a coherent superposition of two locations. At each location, two photon-pair creation can lead to four photons, see Fig. \ref{fig:horizontal_interference}. While the four-photon interference has been observed experimentally \cite{feng2023chip, qian2021multiphoton}, the observation of its non-local nature would require a spatial separation of the four crystals. Our experiment therefore can be seen as a pilot feasibility study for future studies of non-local multiphoton quantum phenomena.

\begin{acknowledgments}
The authors thank Anton Zeilinger for initiating and motivating this research, and Armin Hochreiner and Mayukh Lahiri for valueable discussions on the topic of path identity over the years.
\end{acknowledgments}

\bibliography{apssamp.bib}
\section{Supplementary}
    The photon pairs were created via the SPDC interaction within a nonlinear medium, which was pumped by a coherent laser source.
    The pump light source was a single-mode continuous-wave (cw) laser beam produced in a compact laser diode module manufactured by Ondax with a central wavelength (CWL) of around $405.5\,$nm.
    The pump's spectral distribution had a full-width at half-maximum (FWHM) bandwidth of $\Delta \nu\!=\!160\,$MHz according to the manufacturer, resulting in a coherence time of $t_{coh}^p=2\,$ns and a coherence length of $l_{coh}^p=596\,$mm.

    The maximum power of the pump beam at the location of the first nonlinear medium was measured with a powermeter (PM100A Thorlabs) and yielded $15.40\,$mW$\pm 0.05\,$mW, resulting to an intensity of $7740.88\,$$\mathrm{W} / \mathrm{cm}^{2}$$\pm 25.13\,$ $\mathrm{W} / \mathrm{cm}^{2}$ at the focal point. 
    The above laser intensity and the choice of the properties of the nonlinear crystal (see below) led to a brightness of $B=4.8\times 10^5s^{-1}$, which indicates the photon pairs created at the source per second\footnote{The brightness $B$ was estimated with the single count rates $\mathcal{C}_{A,B}$ and coincident count rate $\mathcal{C}_c$ via $B=\frac{\mathcal{C}_A \cdot \mathcal{C}_B}{\mathcal{C}_c}$, where preliminary background induced counts $\Delta$ and accidental coincident counts $\mathcal{C}_{acc}$ are ignored. Note that these assumptions require measuring the brightness with low ambient light and sufficiently low pump power.}. 
    The laser module was operated at maximum output power, ensuring a strong SPDC signal, but keeping in mind that the impact of higher emission SPDC photons on the visibility is detrimental.
    The pump beam, which is assumed to have a Gaussian beam intensity profile\footnote{A Gaussian beam profile was ensured by coupling the laser beam into a single-mode fiber (Thorlabs S405HP).}, passed a quarter-wave (QWP) and a half-wave plate (HWP) in order to increase the conversion efficiency of the down-conversion beam, as only the projection of linearly polarized light contributes to the conversion efficiency. 
    In the presented case, maximum efficiency was achieved for vertically polarized light ($s$-polarized).

    \begin{figure}[b]
        \centering
        \includegraphics[width=3.51in]{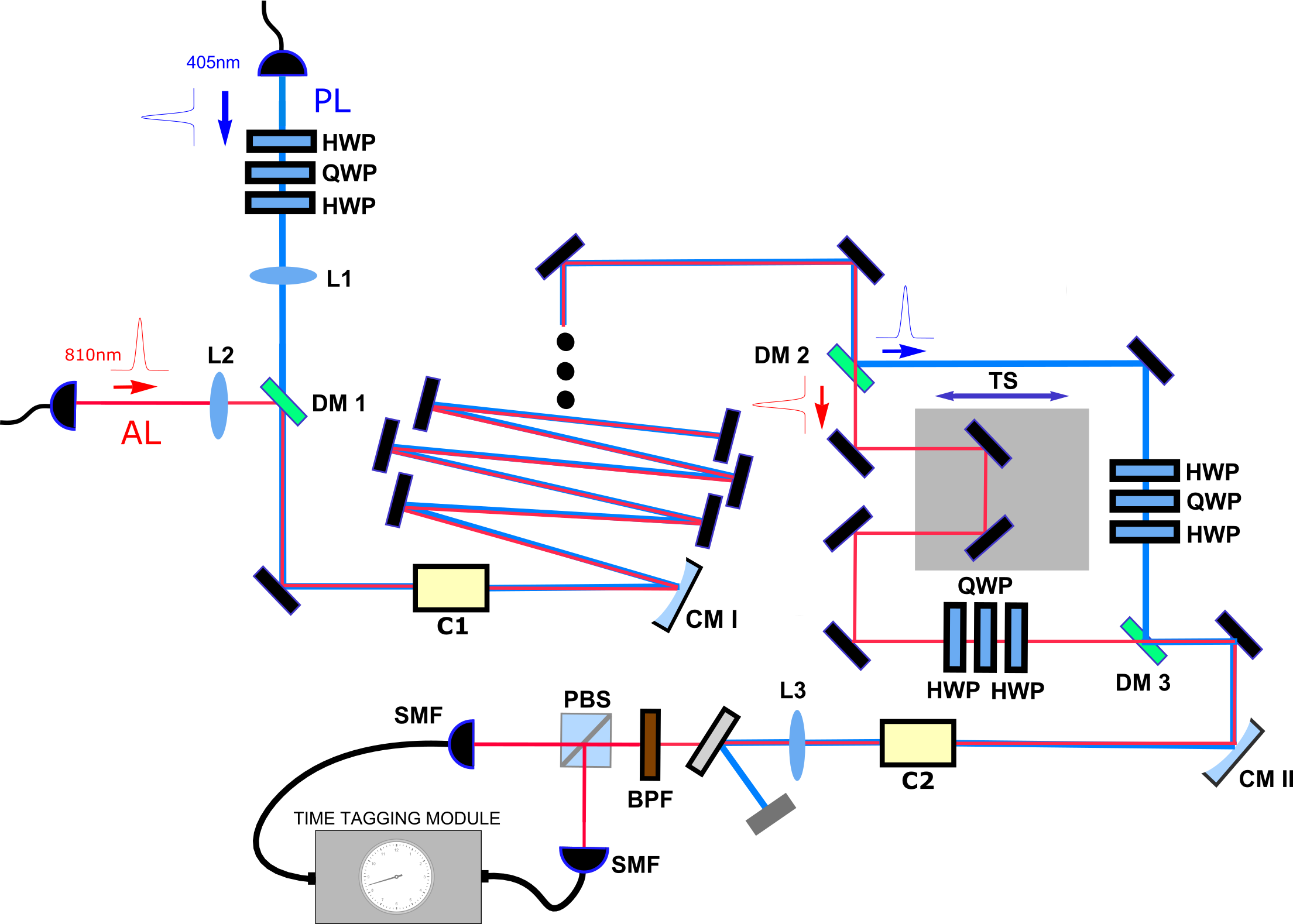}
        \caption{\label{fig:wide} A sketch of the experimental and important points for the alignment is depicted.
        The pump beam is indicated by the blue straight line, the SPDC photons (and the alignment laser) are illustrated in red. 
        The three black dots in the multi-reflection mirror system indicate that more mirrors than depicted were used in the final experiment over longer distances ($<20\,$m). Taken from \cite{pseinerdiss}.}
    \end{figure}
    
    The outcoupling stage of the pump beam comprised of a five-axis single-mode fiber aligner (Newport 9131-M) and a plan achromat objective (Olympus RMS40X), which leads with its optical parameters to a collimated beam radius of $1.55\,$mm.
    Subsequently, the beam was focused into the nonlinear medium with a UV-coated $f=300\,$mm lens, with $f$ being the focal length.
    The focusing condition resulted in a beam waist of $\omega_{focus}^p= 25\,\mu$m, in agreement to the focal parameter of $\xi_p=0.056$, with the crystal length $L=1\,$mm.
    Hence, the Rayleigh length ($z_R$=$\frac{2\pi\lambda}{\omega}$), which determines the length across which the beam can be regarded as a plane wave, was approximately $4.85\,$mm. 
    Therefore, the assumption that the pump beam was a plane wave throughout the $1\,$mm-long medium, was valid.
    To avoid chromatic aberration\footnote{Chromatic aberration describes shift of a lense's focal length depending on the impinging beam's wavelength.} between SPDC ($\lambda_{s/i}=810\,$nm) and pump ($\lambda_p=405\,$nm) photons, two dielectric concave mirrors (Thorlabs CM750-500) with a diameter of $75\,$mm (one each on the sending and receiving site) were implemented.

    Regarding the propagation in free space, a focal length of $f_{CM}\!=\!500\,$mm ensured a Rayleigh length of $51.6\,$m for the pump beam, which suffices in terms of the divergence angle with respect to the aperture diameter of the optical elements ($1^{''}=25.4\,$mm ) used in the setup.
    
    The focal lengths of the two concave mirrors were chosen to be $f_{CM}=500\,$mm throughout all propagation distances.
    This ensured, with the pump waist at the position of the first nonlinear crystal being $\omega_{p}=25\,$ $\!\!\mu\mathrm{m}$ and a resulting collimated beam radius of $\omega^{coll}_p=2.58\,$mm, a Rayleigh length of $51.6\,$m for the pump beam.
    Close-to-optimal coupling efficiency for a given pump beam waist $\omega_p$ was found experimentally for the following relationship between the pump and SPDC focal parameters: $\xi_{s/i}\approx \sqrt{2.84 \xi_p}$ \cite{bennink}.
    Hence, the theoretical SPDC beam waist with $\xi_{s/i}=0.40$ was $\omega_{s/i}=13.6\,$ $\!\!\mu\mathrm{m}$, leading to a collimated beam radius of $\omega^{coll}_{s/i}=9.48\,$mm, resulting in a Rayleigh length of $348.6\,$m.
    In theory, after $70\,$m the beam radii reach  $\omega^{coll70}_{p}=4.35\,$mm and $\omega^{coll70}_{s/i}=9.67\,$mm, respectively.
    Hence, over a maximum distance of $70\,$m, the diffraction limit for the Gaussian-shaped beams does not exceed the diameter of $25.4\,$mm, the maximum aperture diameter of the optical elements used in the setup.
    In reality, the beam divergence exceeds the diffraction limit, which, in combination with misalignment and the used alignment technique for greater distances, could lead to loss of the signal. 
    
    This maximum propagation distance of $70\,$m was chosen in consideration of the finite length of the optical table and the finite number of ultra-broadband coated BK7 mirrors [Semrock MM2-311S-25.4] which were used to let the beams propagate from the sender to receiver. 
    Multiple $25.4\,$mm ultra-broadband mirrors were placed at both ends of the optical table in the lab to let the beams travel from the sending station to the receiving station, while the mirrors were arranged in such a way, that the beam was reflected from one end to the other multiple times.
    In fact, for the maximum propagation distance ($70\,$m), the mirrors were implemented such that double reflections per mirror were possible. 
    To exploit the whole area of the mirror, double reflections per mirror, which are located in two rows on either end of the optical table, were implemented.

    The requirement of the coherence length of the SPDC photons $l_{\text {coh}}^{\text {DC}}$ being much smaller than $\Delta \mathcal{L}$ is ensured automatically, as the idler and signal beams of the two different crystals are within the same mode and hence not manipulated independently (see Fig. \ref{fig:conceptsketch}).
    Hence, under the assumption of the degeneracy of the signal and idler frequencies (hence $\mathrm{k}_{avg}=0$) the coincident count rates with the newly introduced quantities take the following form:
    
    \begin{equation} \label{eq:coinc_rate_end}
        \begin{aligned}
            \mathcal{C}_{c}=|A|^{2} \mathcal{I}_{p} \bar{f}_{s}\bar{f}_{i}[1+\mathrm{V}_{p} \mathrm{V}_{c}\cos \left(\mathrm{k}_{p} \Delta \mathcal{L}+\Delta \Phi \right)].
        \end{aligned}
    \end{equation}

    The biphoton path length difference between down-conversion and pump photons $\Delta \mathcal{L}$ can be changed in the presented experimental arrangement (see Fig. \ref{fig:full_setup}) by splitting the down-conversion fields from the pump field in between the two SPDC sources.
    Then, the path lengths can be changed with an interferometer, where the photons are combined again before propagating through the second source. 
    The non-dynamical phase difference $\Delta \Phi$ includes wavefront distortions acquired through free-space propagation and considers the fact that the group velocities of the pump and the SPDC photons are not equal, leading to walk-off effects, caused by the difference in velocities and directions of energy and phase propagation \cite{spasibko2020spectral,cutipa2022bright}.

\subsection{Supplementary: Results}    
    In the following section, the measurement results of the single and coincident count rates for interfering SPDC photon pairs created in the down-conversion crystals I \& II, with three different distances in between, namely $2\,$, $20$ and $70\,$m, are shown.

    Eq. \ref{eq:coinc_rate_end} predicts an oscillating behavior of the count rates while scanning the phase difference between the SPDC photons created in the first crystal and the pump photons.
    However, the most crucial limitation was distinguishability between the interfering particles, which leads to the fact that the expected visibility was lower than 1.
    Any information that led to the exclusive knowledge about one of the interfering particle's state resulted in a decrease in visibility. 
    Limitations regarding the uncertainty of the visibility were of technical and systematic nature such as shot noise\footnote{Shot noise can be equated to quantum noise and was caused by the intrinsic discreteness and randomness of the SPDC process.} and turbulences in the air which introduced fluctuations to the signal.

    The distributions of visibilities and hence the mean values of $\mathcal{V}$ with respective errors were evaluated by considering the extrema of the overall coincident and single count rates.
    Then, a high number sampling ($\sim\!\!10^5$) with their respective statistical distributions was performed, which were expected to follow a Poissonian distribution. 
    The mean values of the extrema gave the mean value of the visibility $\mathcal{V}$ by simply calculating ($Max$-$Min$)/($Max$+$Min$), where $Max$ ($Min$) denotes the mean value of the maxima (minima) of the oscillating SPDC signal.  
    The extrema for each data set were evaluated by the $FindPeaks$-function of the mathematical processing software \textit{Wolfram Mathematica}\footnote{\textit{Wolfram Mathematica} Version 11.0.1.0}, which returned a list of values of local extrema. 
    The expected oscillation period of the interference fringes within the count rates can be estimated by calculating the expected coincidence count rate of overlapping SPDC signals \cite{jha2008} and considering the motor velocity $v_{m}$ of the trombone system.
    By performing a Monte Carlo simulation via combining the sampled distributions of the extrema of one data set, a high number of visibilities was calculated. 
    
    \subsection{Data analysis for propagation distance of 2m}
    \label{subsec:data1}

    \begin{figure}
        \centering
        \includegraphics[width=3.01in]{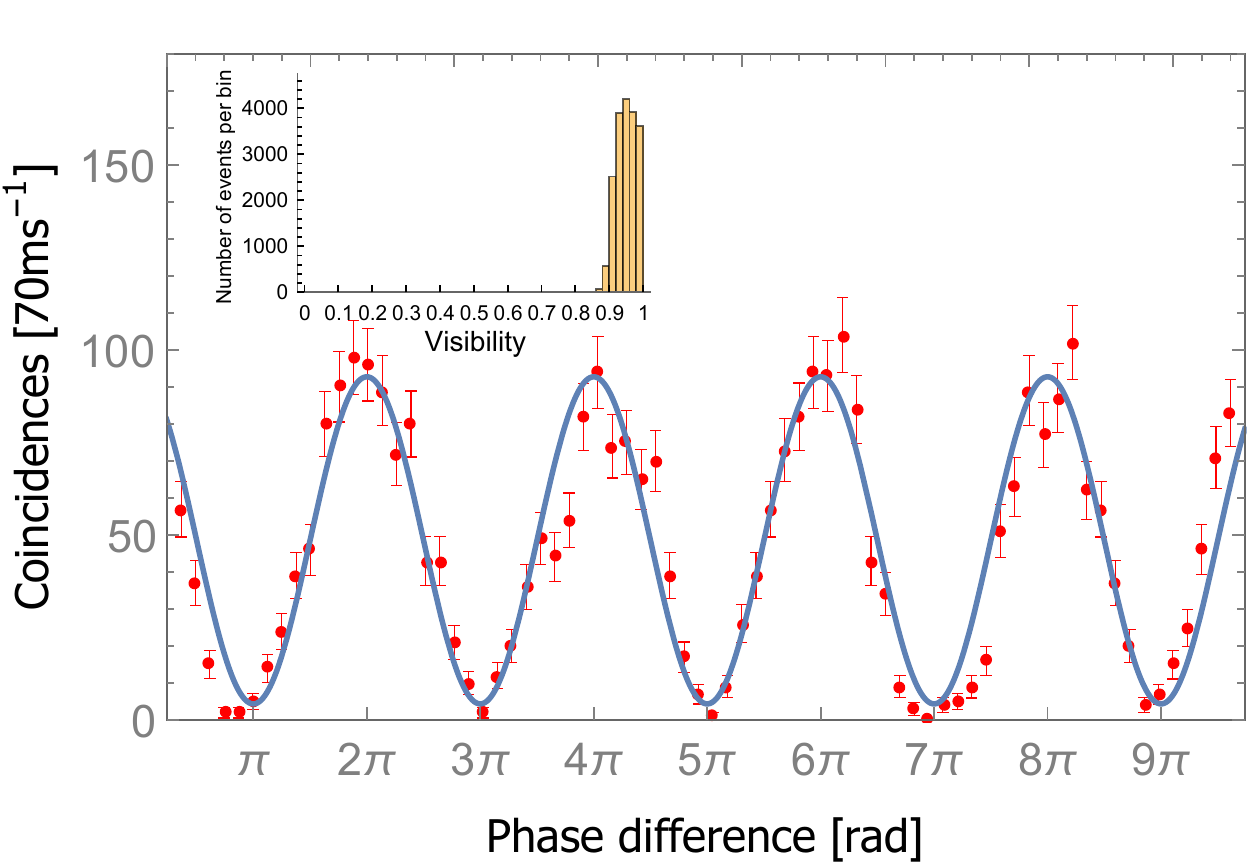}
        \caption{\label{fig:wide}The coincident count rates of SPDC photon pairs created in two coherent nonlinear crystals while moving the trombone system and hence changing $\Delta \mathcal{L}$ after a propagation distance of $2\,$m are shown.
        The red dots represent the experimental results.
        The error bars assume Poissonian distribution.
        The inset shows the visibility distribution calculated by a Monte-Carlo simulation.
        The dots indicate the data points, and the straight lines are fitting curves to the cos-function, consistent with the theoretical prediction following Eq. \ref{eq:coinc_rate_end}.}
        \label{fig:coinc2m}
    \end{figure}
    
    The error bars in the plot of the count rates correspond to counting statistics assuming Poissonian distribution given by $\Delta n=\pm \sqrt{n}$ where $n$ is the number of photons being used in the process of measurement.
    Note that the typical dark count rates for the APDs used in this experiment were not exceeding $2\times10^2$ counts per second (cps), which, compared to the single count rates yielded here, were negligible ($\mathcal{C}_{A,B}\sim10^4\,$cps).
    Moreover, accidental coincident counts $\mathcal{C}_{acc}$, which contributed to the coincidence signal in the form of uncorrelated photon pairs and therefore a non-zero background, could have ultimately led to a decrease in visibility in the coincident count rates.
    The accidental coincident count rates $\mathcal{C}_{acc}=\mathcal{C}_A\mathcal{C}_Bt_c$ were around $0.5\,$cps and $0.036$ counts per integration time $t_{int}=70\,$ms, which, compared to the measurement counts in the order of $50$ counts per $70\,$ms, had negligible impact on the visibility. 
    
    Throughout the measurements with a propagation distance of $2\,$m, the velocity of the stepper motor had been chosen to be $180 \mathrm{nm} / \mathrm{s}$ and the measurements were taken over a time window of $70\,$s. 
    The measured visibilities for the single and coincident count rates were

    \setlength{\abovedisplayskip}{0pt} \setlength{\abovedisplayshortskip}{0pt}
    
    \begin{align*}
        \mathcal{V}_{idler}^{2\mathrm{m}}=19.90\%\pm2.61\%, 
    \end{align*}
    \vspace{-15pt}
    \begin{align*}
        \mathcal{V}_{signal}^{2\mathrm{m}}=18.79\% \pm2.70\%, 
    \end{align*}
    
    \begin{equation}\label{eq:vis2m2}
        \mathcal{V}_{coincidences}^{2\mathrm{m}}=96.15\% \pm2.86\%.
    \end{equation}
    
    As can be seen in the inset of Fig. \ref{fig:coinc2m}, the visibility distribution of the coincident count rates is equipped with an asymmetric error bar, as the value $1$ represents the maximum bound by definition. 
    
    Note that due to the high number of evaluated statistics during measuring ($70\,$s total measurement time per measurement session) and the accumulated high sampling number, the error of the mean value was negligibly small.
    The effects from systematic influences such as phase fluctuations and wavefront distortions due to propagation in free space could be neglected over the distance of $2\,$m.
    This could be extracted from the comparison of the measured and the ideally expected error bars of the visibilities.
    Assuming solely deviations arising from Poissonian distributed shot noise effects (standard deviation equal to $\Delta n=\pm \sqrt{\bar{n}}$, with $\bar{n}$ being the average number of photons) within the count rates would result in following visibility errors by performing Gaussian Error Propagation:
    
    \setlength{\abovedisplayskip}{0pt} \setlength{\abovedisplayshortskip}{0pt}

    \begin{align*}
        \Delta_{sn}\mathcal{V}_{idler}^{2\mathrm{m}}=\pm 1.88\%, 
    \end{align*}
    \vspace{-15pt}    
    \begin{align*}
       \Delta_{sn}\mathcal{V}_{signal}^{2\mathrm{m}}=\pm 2.03\%, 
    \end{align*}
    
    \begin{equation}\label{eq:vis70m2}
       \Delta_{sn}\mathcal{V}_{coincidences}^{2\mathrm{m}}=\pm 2.70\%. 
    \end{equation}
    
    As can be seen, these deviations in visibility compared to the measured values differed only slightly (by the factors 1.39, 1.33, and 1.06) and were expectedly smaller due to imperfections of the laser source and phase fluctuations introduced by the Mach-Zehnder interferometer, hence representing a lower bound of noise also for larger propagation distances, where turbulences in the air will have a significant impact.
    The phase fluctuations arose through mechanical oscillations within the optical elements comprising the interferometer.
    
    \subsection{Data analysis for propagation distance of 20 \& 70m}
    \label{subsec:data2}
    
    \begin{figure}
        \centering
            \includegraphics[width=2.91in]{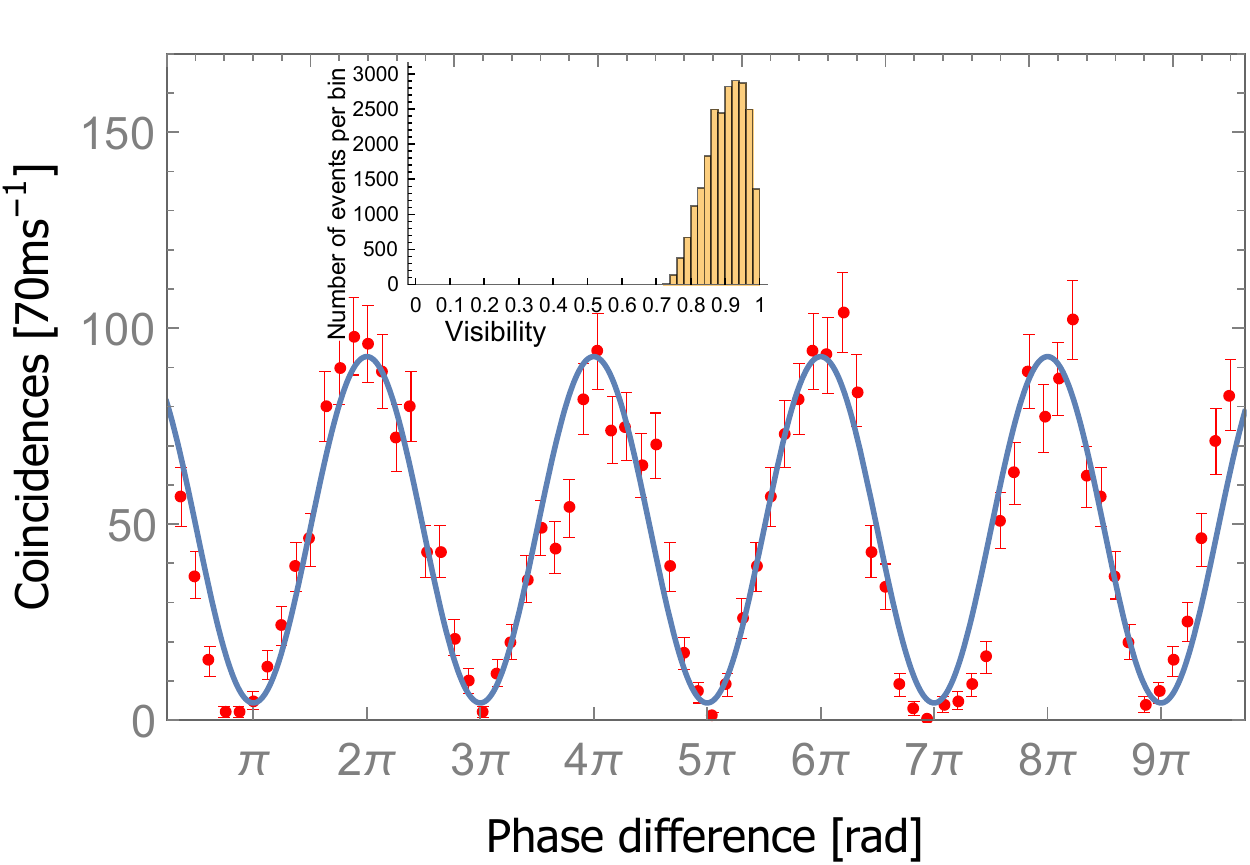}
            \caption{\small{The coincident count rates of down-conversion photons while moving the trombone system and hence changing $\Delta \mathcal{L}$ after a propagation distance of $20\,$m are depicted.
            }}
        \label{fig:coinc20m}
    \end{figure}
    
    By further increasing the propagation distance of the beams between the two nonlinear crystals, additional challenges arose. 
    As discussed in section \ref{sec:methods}, multiple 1’’ ultra-broadband BK7 mirrors were placed at both ends of the optical table to let the beams travel from the sending station to the receiving station. 

    For the measurements the velocity of the stepper motor $v_m$ had been chosen to be, again, $180\,$nm/s and the integration time $t_{int}$ of the detection system $70\,$ms.
    The measured visibilities for the single and coincident count rates were
    
    \setlength{\abovedisplayskip}{0pt} \setlength{\abovedisplayshortskip}{0pt}

    \begin{align*}
        \mathcal{V}_{idler}^{20\mathrm{m}}=26.16\%\pm7.32\%, 
    \end{align*}
    \vspace{-15pt}
    \begin{align*}
        \mathcal{V}_{signal}^{20\mathrm{m}}=35.62\% \pm7.08\%, 
    \end{align*}
    
    \begin{equation}\label{eq:vis20m2}
        \mathcal{V}_{coincidences}^{20\mathrm{m}}=92.05\%\pm5.65\%.
    \end{equation}

    The post-processing approach for the shown results (also for $70\,$m) was equal to the approach presented for the measurements over $2\,$m. 
    By comparing these results of $\mathcal{V}^{20\mathrm{m}}$ with the ones over $2\,$m ($\mathcal{V}^{2\mathrm{m}}$) one notices only a slight decrease in visibility regarding the coincident count rates, in combination with an increase in error bars in the coincident count rates. 
    
    The decrease in visibility, however, was attributed to the effects of beam wandering through the longer free-space propagation in turbulent air, and hence intensity matching turned out to be experimentally more challenging.
    The visibility therefore would be expected to decrease as the signal detectors “see” more non-interfering background \cite{herzog1994frustrated}.
    Apparently, this effects also explained the deviations from the mean value of the visibility distribution, which was emphasized by comparing them to the "shot-noise" induced deviations:
    
    \setlength{\abovedisplayskip}{0pt} \setlength{\abovedisplayshortskip}{0pt}

    \begin{align*}
        \Delta_{sn}\mathcal{V}_{idler}^{20\mathrm{m}}=\pm 1.97\%, 
    \end{align*}
    \vspace{-15pt}    
    \begin{align*}
       \Delta_{sn}\mathcal{V}_{signal}^{20\mathrm{m}}=\pm 1.99\%, 
    \end{align*}
    
    \begin{equation}\label{eq:vis70m2}
       \Delta_{sn}\mathcal{V}_{coincidences}^{20\mathrm{m}}=\pm 4.97\%.
    \end{equation}
     
    A significant increase of the deviation in visibility could not be observed within the coincident count rates (factor 1.17).
    Moreover, in contrast to the $2\,$m measurements, the errors within the coincident count rates were smaller than the ones for single count rates.
    This fact came from the intrinsic property of discrete event counting statistics such as photons created in an intrinsically random SPDC process.
    In quantum optics, counting photons enables detecting and identifying quantum states. 
    As observed over $2\,$m, the low photon number of the coincident count rates ($\sim \!10^2$ per integration time $t_{int}$) broadens the error bar significantly.
    Further quantitative statements regarding the impact of additional noise such as induced by turbulences in the atmosphere can solely be done by the number of photons $n$ while comparing different measurements (such as $2\,$ with $20\,$m) with detecting equal intensities.
    
    \begin{figure}
        \centering
            \includegraphics[width=2.91in]{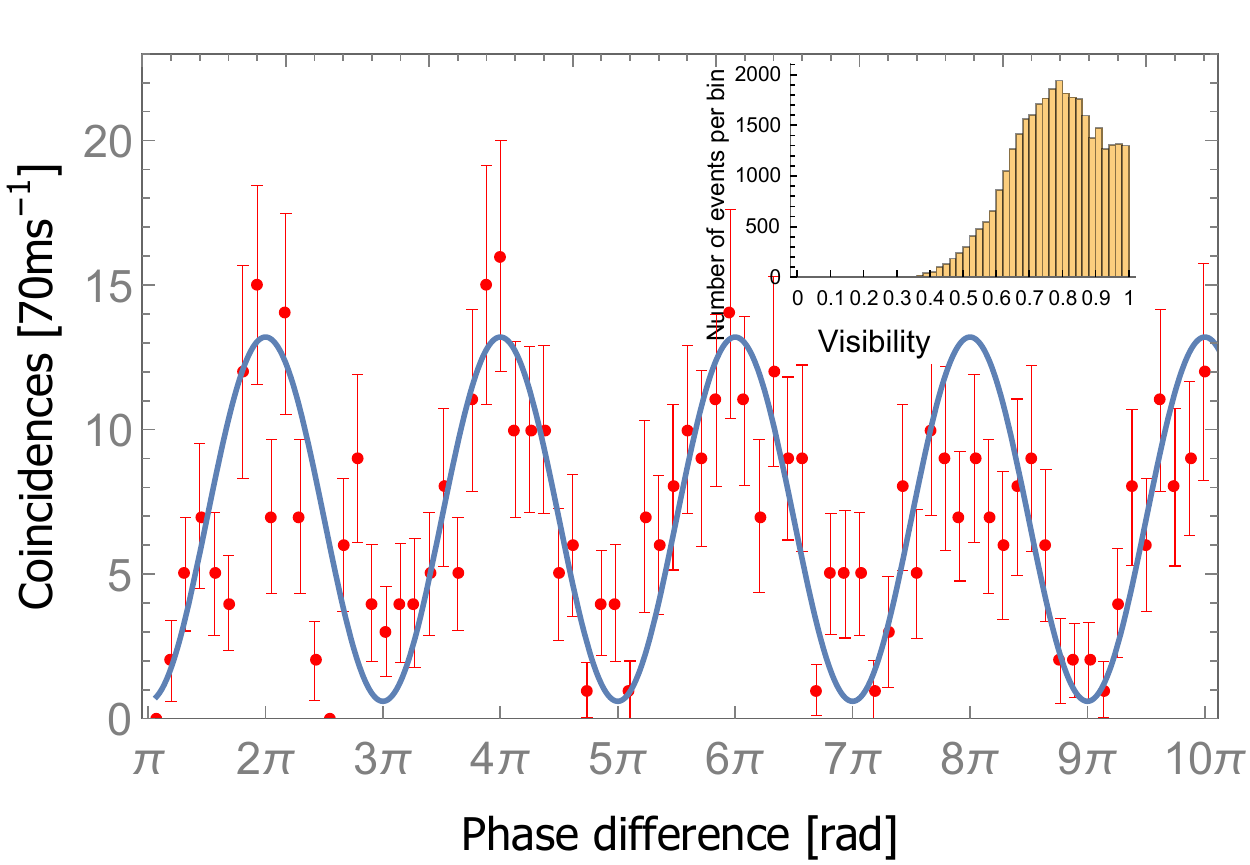}
            \caption{\small{The coincident count rates of down-conversion photons while moving the trombone system and hence changing $\Delta \mathcal{L}$ after a propagation distance of $70\,$m are depicted.
            }}
        \label{fig:coinc70m}
    \end{figure}
    
    Figure \ref{fig:coinc70m} depicts the results of the count rate measurements over a traveling distance of $70\,$m between the two down-conversion sources.
    The integration time $t_{int}$ and the stepper motor velocity $v_m$ were chosen to be the same as for the measurements for $2\,$m and $20\,$m. 
    The results reflect, that the visibilities decreased and reached the following values for coincident and for single count rates:
    
    \setlength{\abovedisplayskip}{0pt} \setlength{\abovedisplayshortskip}{0pt}

    \begin{align*}
        \mathcal{V}_{idler}^{70m}=8.42\%\pm 5.20\%, 
    \end{align*}
    \vspace{-15pt}    
    \begin{align*}
        \mathcal{V}_{signal}^{70m}=10.86\% \pm 3.44\%, 
    \end{align*}
    
    \begin{equation}\label{eq:vis70m2}
        \mathcal{V}_{coincidences}^{70m}=83.90\%\pm 12.98\%. 
    \end{equation}

    The mean values of the visibilities $\mathcal{V}^{70\mathrm{m}}$ compared to $20\,$m further decreased but accompanied by decreasing errors for the signal and idler count rates.
    The visibility distribution within the coincident count rates showed a distribution around the mean value of $83.9\%$, as during the alignment process the visibilities were optimized for the coincidences only.
    Apparently, in contrast to the single count rates, the errors of the visibility for the coincident count rates increased, significantly.
    
    The expected errors arising solely from shot noise were once more estimated via Gaussian error propagation:
    
    \setlength{\abovedisplayskip}{0pt} \setlength{\abovedisplayshortskip}{0pt}

    \begin{align*}
         \Delta_{sn}\mathcal{V}_{idler}^{70\mathrm{m}}=\pm 2.41\%, 
    \end{align*}
    \vspace{-15pt}    
    \begin{align*}
       \Delta_{sn}\mathcal{V}_{signal}^{70\mathrm{m}}=\pm 3.17\%,
    \end{align*}
    
    \begin{equation}\label{eq:vis70m2}
       \Delta_{sn}\mathcal{V}_{coincidences}^{70\mathrm{m}}=\pm 14.20\%.
    \end{equation}
    
    Unexpectedly, with respect to the measured errors and comparing them to the results over $20\,$m, no further increase of the errors could be observed (factors for singles: 2.16 and 1.09, coincidences: 0.91), as one would have expected due to the larger propagation distance. 
    Note,that a value smaller than 1 signifies that the measured standard deviation is smaller than the expected standard deviation arising from the shot noise. 
    This artificial fact could solely be attributed to the imperfect post-processing method and could be neglected as long the value differs only slightly from 1.
    The reason could be found in the photon count rates $n$, as within discrete events following Poissonian distribution, the relative error decreased for high $n$.
    The mean photon number over the $70\,$m measurements for the coincident count rates (few counts per integration time $t_{int}$) as well as the single count rates ($<10^3$ counts per $t_{int}$) were significantly lower than compared to $2\,$ and $20\,$m (coincidences: $\sim \!10^2$ counts per $t_{int}$, singles $\sim \!1.5\times 10^3$ counts per $t_{int}$).
    
    Compared to the errors in the measured visibility distributions over $20\,$m, the decrease in error could be explained by the dominant shot noise error due to the low photon number.
    Finally, these significant fluctuations in intensity due to the small photon number $n$ resulted in an experimentally challenging matching of the SPDC intensities, which in the end represented the main cause for the low visibilities.
\end{document}